\documentclass{article}

\usepackage{arxiv}

\usepackage[utf8]{inputenc} % allow utf-8 input
\usepackage[T1]{fontenc}    % use 8-bit T1 fonts
\usepackage{hyperref}       % hyperlinks
\usepackage{url}            % simple URL typesetting
\usepackage{booktabs}       % professional-quality tables
\usepackage{amsfonts}       % blackboard math symbols
\usepackage{nicefrac}       % compact symbols for 1/2, etc.
\usepackage{microtype}      % microtypography
\usepackage{lipsum}
\usepackage{cite}
\usepackage{amsmath}
\usepackage{graphicx}

\title{Non-proportional hazards in immuno-oncology: is an old perspective needed?}

\author{
  Dominic Magirr \\
  Advanced Methodology and Data Science\\
  Novartis Pharma AG\\
  Basel, Switzerland \\
  \texttt{dominic.magirr@novartis.com} \\
}

\begin{document}
\maketitle

\begin{abstract}
A fundamental concept in two-arm non-parametric survival analysis is the comparison of observed versus expected numbers of events on one of the treatment arms (the choice of which arm is arbitrary), where the expectation is taken assuming that the true survival curves in the two arms are identical. This concept is at the heart of the counting-process theory that provides a rigorous basis for methods such as the log-rank test. It is natural, therefore, to maintain this perspective when extending the log-rank test to deal with non-proportional hazards, for example by considering a weighted sum of the "observed - expected" terms, where larger weights are given to time periods where the hazard ratio is expected to favour the experimental treatment. In doing so, however, one may stumble across some rather subtle issues, related to the difficulty in ascribing a causal interpretation to hazard ratios, that may lead to strange conclusions. An alternative approach is to view non-parametric survival comparisons as permutation tests. With this perspective, one can easily improve on the efficiency of the log-rank test, whilst thoroughly controlling the false positive rate. In particular, for the field of immuno-oncology, where researchers often anticipate a delayed treatment effect, sample sizes could be substantially reduced without loss of power.
\end{abstract}

\section{Introduction}

The emergence of checkpoint inhibitors in immuno-oncology has provoked much discussion about whether the log-rank test (or Cox model) should continue to be used as a default primary analysis method in confirmatory studies, or whether it is acceptable to use a method more tailored to non-proportional hazards \cite{freidlin2019methods, uno2020log, huang2020estimating, freidlin2020reply}. The source of the controversy is the clear tendency, when this kind of drug is compared with chemotherapy, for survival curves to remain more-or-less equal for a number of months before they diverge. The log-rank test, while valid in terms of controlling the false positive rate, may lose power under this scenario, and alternative methods, it is argued, can better distinguish whether an experimental treatment is efficacious or not.  

As a helpful guide for this discussion, multiple simulation studies have been completed that compare a wide array of potential methods under a range of possible departures from proportional hazards \cite{lin2020alternative, royston2020simulation, chen2020comparison}. The methods generally fall into two categories: those derived from the Kaplan-Meier curves \cite{royston2013restricted, tian2017efficiency, uno2014moving, pepe1989weighted}, and modifications to the log-rank test \cite{gares2014comparison, karrison2016versatile, yang2010improved, harrington1982class}. Scenarios extend well beyond the delayed-effect case just discussed, and also cover diminishing effects, crossing survival curves, and cure-rate models. The favoured approaches emanating from this work \cite{roychoudhury2019robust} appear to be max-z-type statistics \cite{karrison2016versatile} based on multiple weighted log-rank tests from the Fleming-Harrington-($\rho$, $\gamma$) family \cite{harrington1982class}. However, the response has not been universally positive. In particular, Freidlin \& Korn \cite{freidlin2019methods} caution that such a test is not ready to be used as the primary analysis for a confirmatory study. As part of their argument, they construct a scenario where the experimental treatment is uniformly worse than control, yet the test would have a high chance of claiming a positive result.

The issue highlighted by Freidlin \& Korn is partly a consequence of the way that we tend to view weighted log-rank tests. Namely, as a weighted sum of observed versus expected numbers of events on one of the treatment arms (the choice of which arm is arbitrary), where the expectation is taken assuming that the true survival curves in the two arms are identical. The purpose of this article is to demonstrate that when one switches perspective, and instead views a weighted log-rank test as a permutation test \cite{lan1990linear, leton2001equivalence}, the deficiencies of certain choices of weighting become obvious, and so does the way to fix it. Once this change of perspective has been achieved, the article will finish with a thorough discussion of related issues: sample size calculation, treatment effect estimation, and interpretation of results.

\section{"Observed vs. Expected" tests}
\label{sec:ove}

\subsection{Mantel's test (1966)}

What we now know as the "log-rank" test was introduced by Mantel \cite{mantel1966evaluation}. Somewhat confusingly, its derivation has nothing apparent to do with logarithms, nor ranks. Consider the toy data set in Table \ref{toy_data} (which we will use throughout).

\begin{table}[ht]
\caption{A toy data set, consisting of observations $x$ and treatment labels $z$. A "+" suffix indicates a censored observation.}
\centering
\begin{tabular}{cc}
  \hline
$x$ & $z$ \\ 
  \hline
2 & 0 \\ 
  6+ & 0 \\ 
  7 & 1 \\ 
  8 & 0 \\ 
  9+ & 1 \\ 
  11 & 0 \\ 
  13 & 1 \\ 
  17 & 0 \\ 
  22 & 1 \\ 
  23 & 1 \\ 
  24+ & 0 \\ 
  30 & 1 \\ 
   \hline
\end{tabular}
\label{toy_data}
\end{table}

The idea is to go through the ordered distinct event times $t_j$ ($j=1,\ldots,k$), constructing $k$ 2x2 tables, such as Table \ref{two_by_two}.

\begin{table}[ht]
\caption{A 2x2 table describing the situation at event time $t_j$.}
\centering
\begin{tabular}{c|cc|c}
  & Event = Y & Event = N &   \\ 
  \hline
Trt = 1 & d\textsubscript{1,j} & n\textsubscript{1,j} - d\textsubscript{1,j} & n\textsubscript{1,j} \\ 
  Trt = 0 & d\textsubscript{0,j} & n\textsubscript{0,j} - d\textsubscript{0,j} & n\textsubscript{0,j} \\ 
   \hline
  & d\textsubscript{j} & n\textsubscript{j} - d\textsubscript{j} & n\textsubscript{j} \\ 
  \end{tabular}
  \label{two_by_two}
\end{table}

Conditional on the margins of each 2x2 table, and assuming identical survival curves, the observed number of events on the experimental treatment at event time $t_j$, denoted by $O_{1,j}$, follows a hypergeometric distribution, where the expected number of events is $E_{1,j} = d_j\times n_{1j} / n_j$, and the variance of $O_{1,j}$ is 

\begin{equation*}
V_{1,j} = \frac{n_{0,j}n_{1,j}d_j(n_j - d_j)}{n_j^2(n_j - 1)}.
\end{equation*}

In our example, at the first event time, $O_{1,1} = 0$ and $E_{1,1} = 1 \times 6 / 12 = 0.5$. Continuing in this way we would produce Table \ref{oev}. Unconditionally, although the terms $O_{1,j} - E_{1,j}$ are not independent, they are uncorrelated with mean 0 and variance $V_{1,j}$ \cite[p.~16]{proschan2006statistical}. It can be shown that, asymptotically,
\begin{equation}\label{U}
U := \sum_{j} \left( O_{1,j} - E_{1,j}\right) \sim N(0, \sum_jV_{1,j}).
\end{equation}
If the experimental treatment is beneficial, then the observed number of events on the experimental arm will tend to be lower than what would be expected under an assumption of identical survival curves. One is hoping, therefore, to see that $U<<0$ and that the one-sided p-value, $p:=\Phi(U / \sqrt{\text{var}(U)})$, is less than, e.g., $\alpha = 0.025$. The sample size in our toy example is too small to expect the normal approximation to be adequate. Nevertheless, for illustration purposes, we would have $U = -0.91$, $\text{var}(U)= 1.85$ and $p = 0.25$.

\begin{table}[ht]
\caption{Observed and expected numbers of events on the experimental arm at each event time for the toy data set in Table \ref{toy_data}}
\centering
\begin{tabular}{cccc}
  \hline
t\textsubscript{j} & O\textsubscript{1,j} & E\textsubscript{1,j} & V\textsubscript{1,j} \\ 
  \hline
2 & 0 & 0.50 & 0.25 \\ 
  7 & 1 & 0.60 & 0.24 \\ 
  8 & 0 & 0.56 & 0.25 \\ 
  11 & 0 & 0.57 & 0.24 \\ 
  13 & 1 & 0.67 & 0.22 \\ 
  17 & 0 & 0.60 & 0.24 \\ 
  22 & 1 & 0.75 & 0.19 \\ 
  23 & 1 & 0.67 & 0.22 \\ 
  30 & 1 & 1.00 & 0.00 \\ 
   \hline
\end{tabular}
\label{oev}
\end{table}

\subsection{Weighted log-rank tests}

The form of the log-rank statistic (\ref{U}), a sum of observed minus expected events, is straightforward and intuitive. But why weight each event time equally when we anticipate a delayed treatment effect? It should be possible to find a more powerful test by replacing $U$ with a weighted sum of observed minus expected events,
\begin{equation*}
    U_W := \sum_{j} w_j\left( O_{1,j} - E_{1,j}\right)\sim N(0, \sum_jw_j^2V_{1,j}).
\end{equation*}
The Fleming-Harrington-(0,1) test is a popular choice, where $w_j = 1 - \hat{S}(t_j-)$, i.e., one minus the Kaplan-Meier estimate of the survival probability just before time $t_j$, based on the pooled data from both treatment arms. The weight for the first event is exactly $0$, with the weights increasing towards $1$. The intuition behind the test remains the same. If the treatment is beneficial, we will tend to see fewer events on the treatment arm than would be expected assuming the curves are identical. Now, however, we anticipate seeing $O_{1,j} < E_{1,j}$ more frequently at later timepoints than at earlier timepoints, and are upweighting these later events accordingly. We are hoping to increase the likelihood that $U_W << 0$ and that $p_W < \alpha$, where $p_W:= \Phi(U_W / \sqrt{\text{var}(U_W)})$.

\subsection{Issues}

\begin{figure}
  \centering
  \includegraphics[width=\textwidth]{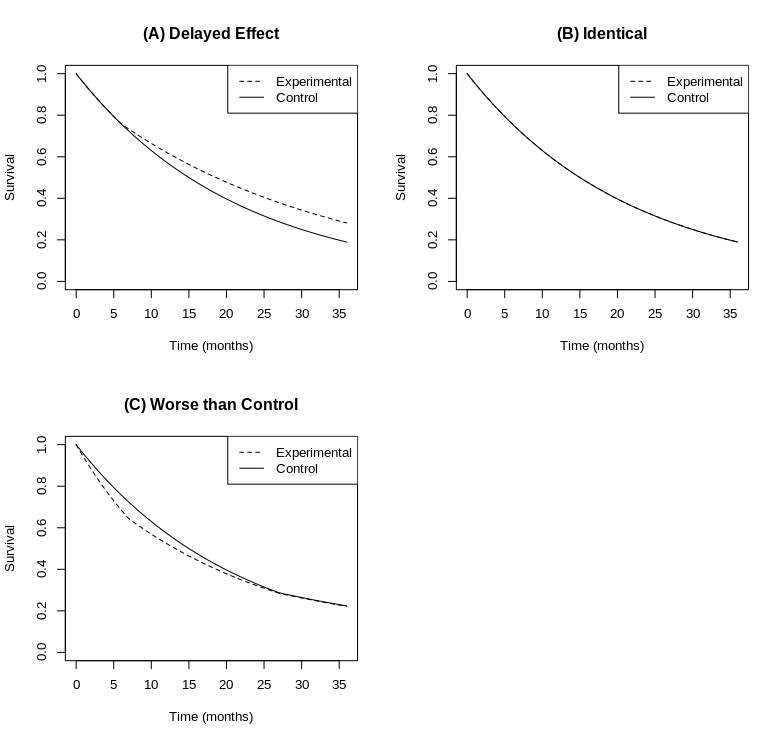}
  \caption{Scenarios used in the simulation study. (A) Control arm: exponential with median $15$ months. Experimental arm: two-piece exponential with rate $\log(2)/15$ during months $0$ to $6$, and $\log(2)/21$ thereafter. (B) Both arms exponential with median $15$ months. (C) Control arm: two-piece exponential with changepoint at $27$ months with rates $\log(2) / 15$ and $\log(2) / 25$.  Experimental arm: three-piece exponential with changepoints at $7$ and $27$ months, and event rates $\log(2)/11$, $\log(2)/17$ and $\log(2)/25$.}
  \label{sim_scenarios}
\end{figure}

To see the potential benefits, but also problems, with using the Fleming-Harrington-(0,1) test, a small simulation study is helpful. Consider a randomized controlled trial where 500 patients are recruited to each treatment arm uniformly over a 12 month period. Three scenarios are considered, described in Figure \ref{sim_scenarios}. Suppose that the trial ends 36 months after the start of recruitment, at which point the event times of all surviving patients are censored. Results are shown in Table \ref{sim_1}.

\begin{table}[ht]
\caption{Results from the simulation study described in the text and Figure \ref{sim_scenarios}. Based on 1000 repetitions.}
\centering
\begin{tabular}{c|cc}
  & \multicolumn{2}{c}{Probability of claiming benefit} \\
Scenario & Log-rank & Fleming-Harrington-(0,1) \\ 
  \hline
(A) Delayed Effect & 0.83 & 0.93 \\ 
  (B) Identical & 0.02 & 0.03 \\ 
  (C) Worse than Control & 0.00 & 0.07 \\ 
   \hline
\end{tabular}
\label{sim_1}
\end{table}

For the delayed treatment effect scenario, the Fleming-Harrington-(0,1) test has much higher power than the log-rank test. To put this difference in context, its relative efficiency is
\begin{equation*}
    100\times\left\lbrace \frac{\Phi^{-1}(0.975) + \Phi^{-1}(0.93)}{\Phi^{-1}(0.975) + \Phi^{-1}(0.83)} \right\rbrace^2 = 139\%.
\end{equation*}
This means that to achieve the same power as the Fleming-Harrington-(0,1) test, the sample size for the log-rank test would need to be increased by about 40\%. As dictated by the theory, both tests control the type 1 error rate when survival curves are identical. The interesting scenario is (C), where the survival probability on the experimental arm is lower than control at all timepoints. Intuitively, one might anticipate a very low chance of claiming $p_W < \alpha$. This is indeed the case for the log-rank test, but for the Fleming-Harrington-(0,1) test we see that this probability is 7\%. The explanation is that despite a uniformly lower survival function, the hazard functions in this example are actually crossing, and (weighted) log-rank tests are essentially a comparison of hazard functions. The Fleming-Harrington-(0,1) test gives low weight to early timepoints when the hazard is higher on the experimental arm than control, and high weight to later timepoints when the hazard ratio favours the experimental arm. This observation is not new. It appears in Fleming \& Harrington's book \cite[p.~267]{fleming2011counting}. More recently, Freidlin \& Korn use this property to argue that the Fleming-Harrington-(0,1) test is unsuitable for regulatory decision making. The idea that a comparison of hazard functions might appear to favour an experimental treatment, despite there being no benefit, has been widely discussed from an estimation and causal inference perspective \cite{hernan2010hazards, aalen2015does, bartlett2020hazards}.

\section{Permutation tests}

\subsection{Wilcoxon's rank-sum test (1945)}

An alternative way to perform non-parametric survival comparisons is via permutation tests, building on Wilcoxon's (1945) rank-sum test \cite{wilcoxon1992individual}. Consider the data in Table \ref{wilcoxon}, where (for the moment) the issue of censoring is ignored, and we assume all values of $x$ correspond to observed events. The idea is to assign a "score", denoted by $a$, to each observation. In this case, we have given the longest survival time a score of 1, the second longest a score of 2, and so on. The test statistic is simply the mean score on the experimental treatment minus the mean score on the control treatment. To perform inference, one considers the data $x$ as fixed, and the treatment labels $z$ as random. In a randomized controlled trial it is often reasonable to consider each possible permutation of the treatment labels as equally likely (for more tailored randomization tests see, e.g., \cite{proschan2019re}). Therefore, one permutes the $z$ very many times, each time re-calculating the test statistic, as is shown in Table \ref{wilcoxon}. The p-value is the proportion of times that the permutation test statistic is lower than or equal to the observed test statistic. In this case, $\bar{a}_1 - \bar{a}_0=-1.16$. Considering all treatment assignments as equally likely, the proportion of times the permutation test statistic is less than or equal to $-1.16$ is about 0.15.

\begin{table}[ht]
\caption{How to perform a Wilcoxon rank-sum test when there are no censored observations.}
\centering
\begin{tabular}{cccc|cccc}
  \hline
 & $x$ & $a$ & $z$ & $z_1^*$ & $z_2^*$ & ... & $z_p^*$ \\ 
  \hline
 & 2 & 12 & 0 & 0 & 0 &  & 1 \\ 
   & 6 & 11 & 0 & 0 & 1 &  & 0 \\ 
   & 7 & 10 & 1 & 0 & 0 &  & 0 \\ 
   & 8 & 9 & 0 & 1 & 0 &  & 0 \\ 
   & 9 & 8 & 1 & 1 & 1 &  & 1 \\ 
   & 11 & 7 & 0 & 1 & 0 &  & 1 \\ 
   & 13 & 6 & 1 & 0 & 1 &  & 1 \\ 
   & 17 & 5 & 0 & 1 & 0 &  & 1 \\ 
   & 22 & 4 & 1 & 0 & 1 &  & 0 \\ 
   & 23 & 3 & 1 & 0 & 1 &  & 0 \\ 
   & 24 & 2 & 0 & 1 & 0 &  & 0 \\ 
   & 30 & 1 & 1 & 1 & 1 &  & 1 \\ 
   \hline
$\bar{a}_1 - \bar{a}_0$ &  &  & -1.16 & -1.16 & -1 &  & 0 \\ 
   \hline
\end{tabular}
\label{wilcoxon}
\end{table}

\subsection{Gehan's test (1965)}

\begin{figure}
  \centering
  \includegraphics[width=\textwidth]{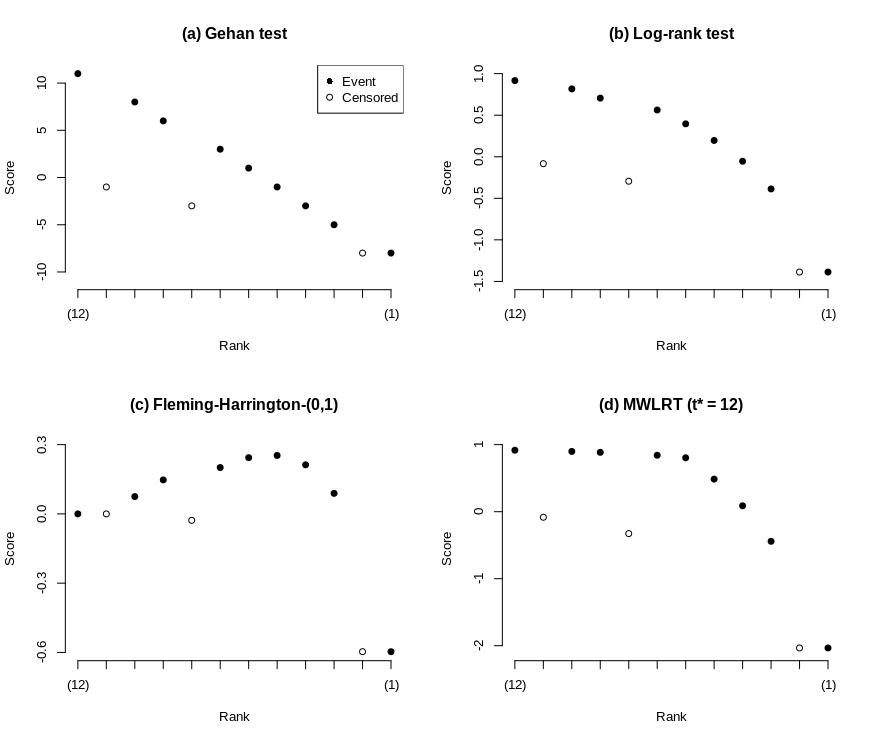}
  \caption{Scores from four permutation-of-scores tests applied to the toy data set from Table \ref{toy_data}. To derive the ranks, observations are ordered regardless of censoring. The observation corresponding to the largest $x$ gets rank $(1)$, etc.}
  \label{scores}
\end{figure}

How to extend the Wilcoxon rank-sum test to deal with censored observations? This question was tackled by Gehan \cite{gehan1965generalized}. His proposal is simple and intuitively appealing. The idea, as shown in Table \ref{gehan}, is to first order the pooled data, including censored observations. Then, for each observation, count the number of patients that definitely have a longer survival time, as well as the number that definitely have a shorter survival time. The score assigned to each observation is simply the difference: \#"better" - \#"worse". The resulting scores are plotted in Figure \ref{scores}(a). One can see that the scores remain essentially a linear function of the ranks, where longer survival times get lower scores, and censored observations "slot in". For example, an early censored observation gets the same score as an observed event somewhere towards the middle of the distribution. The test procedure then proceeds in exactly the same way as for the Wilcoxon rank-sum test. For our example, the difference in the average score (experimental - control) is -1.67. If one were to permute the treatment labels very many times, recalculating the difference in average score each time, one would find that the proportion of times that this difference is less than or equal to -1.67 is about 0.19. This is the p-value from the Gehan test.

\begin{table}[ht]
\caption{How to perform Gehan's (1965) test for the toy data set from Table \ref{toy_data}.}
\centering
\begin{tabular}{cc|ccc|c|cccc}
  \hline
 & $x$ & \#"Better" & \#"Worse" & $a$:= \#"Better" - \#"Worse" & $z$ & $z_1^*$ & $z_2^*$ & ... & $z_p^*$ \\ 
  \hline
 & 2 & 11 & 0 & 11 & 0 & 0 & 0 &  & 1 \\ 
   & 6+ & 0 & 1 & -1 & 0 & 0 & 1 &  & 0 \\ 
   & 7 & 9 & 1 & 8 & 1 & 0 & 0 &  & 0 \\ 
   & 8 & 8 & 2 & 6 & 0 & 1 & 0 &  & 0 \\ 
   & 9+ & 0 & 3 & -3 & 1 & 1 & 1 &  & 1 \\ 
   & 11 & 6 & 3 & 3 & 0 & 1 & 0 &  & 1 \\ 
   & 13 & 5 & 4 & 1 & 1 & 0 & 1 &  & 1 \\ 
   & 17 & 4 & 5 & -1 & 0 & 1 & 0 &  & 1 \\ 
   & 22 & 3 & 6 & -3 & 1 & 0 & 1 &  & 0 \\ 
   & 23 & 2 & 7 & -5 & 1 & 0 & 1 &  & 0 \\ 
   & 24+ & 0 & 8 & -8 & 0 & 1 & 0 &  & 0 \\ 
   & 30 & 0 & 8 & -8 & 1 & 1 & 1 &  & 1 \\ 
   \hline
$\bar{a}_1 - \bar{a}_0$ &  &  &  &  & -1.67 & -1.83 & -3.17 &  & 0.5 \\ 
   \hline
\end{tabular}
\label{gehan}
\end{table}

\subsection{Log-rank test}

Mantel's (1966) test can also be expressed as a permutation test based on scores. The only difference is the way that the scores, $a$, are assigned to the observations. A  detailed derivation is provided in the Appendix. To summarise, one first constructs the Nelson-Aalen estimate, $\bar{S}(t)$, of the survival curve based on the pooled data from both arms. An observation that is censored at time $x_i$ receives a score of $a=\log\bar{S}(x_i)$. An observed event at time $x_i$ receives a score of $a=1 + \log\bar{S}(x_i)$. For our toy example, the resulting scores are shown in Figure \ref{scores}(b). It is now immediately apparent why it is called the "log-rank" test: the scores are a logarithmic function of the ranks. Inference follows the same method as before.  The test statistic is the mean score on the experimental arm minus the mean score on the control arm. One then permutes the treatment labels very many times, each time calculating the difference in average scores, to see how often this is less than or equal to the observed test statistic. In our case the p-value is 0.26. This closely matches the p-value found in Section \ref{sec:ove}. 

\subsection{Optimal permutation-of-scores tests}

We have just considered two different ways of assigning scores to (potentially censored) observations when comparing survival on two treatment arms via a permutation test. One could, of course, think of many more ways to assign such scores. Peto \& Peto \cite{peto1972asymptotically} showed that the log-rank scores are optimal under proportional hazards. Their proof can be sketched out in three steps:

1. Firstly, they show that if one has independent observations $x_1,\ldots,x_n$  from any well-behaved parametric model $f(\theta)$, with $\theta = \theta_0$ for treatment $0$, and $\theta = \theta_1$ for treatment $1$, then an asymptotically efficient test statistic for $H_0:\theta_0 = \theta_1$ is 
\begin{equation}\label{peto_test}
    \sum_{i=1}^n \mathbb{I}\left\lbrace \text{patient } i \text{ on trt }1 \right\rbrace \times \left.\frac{\partial l_i(\theta)}{\partial \theta}\right\rvert_{\theta=\hat{\theta}}
\end{equation}
i.e., one considers the derivative of the log-likelihood contributions evaluated at the maximum likelihood estimate (assuming there is a common parameter $\theta$ across both arms), and takes the sum of these contributions over one arm only.

2. Next they ask: what does proportional hazards look like for a parametric model? Using the well-known relationship between the survival function and hazard function, $S(t)=\exp\left(\int_0^t h(u)du\right)$, one can see that $S(t;\theta):= G(t)^\theta$, where $G$ is an arbitrary basis survival curve, represents a proportional-hazards model. In this case,
\begin{equation}\label{score_obs}
    \left.\frac{\partial l_i(\theta)}{\partial \theta}\right\rvert_{\theta=\hat{\theta}} = \frac{1}{\hat{\theta}} \left[ 1 + \log\left\lbrace G(x_i)^{\hat{\theta}} \right\rbrace \right]
\end{equation}
when $x_i$ corresponds to an observed event, and
\begin{equation}\label{score_cen}
    \left.\frac{\partial l_i(\theta)}{\partial \theta}\right\rvert_{\theta=\hat{\theta}} = \frac{1}{\hat{\theta}} \log\left\lbrace G(x_i)^{\hat{\theta}} \right\rbrace
\end{equation}
when $x_i$ corresponds to a censored event.

3. Finally, they replace the parametric maximum likelihood estimator $S(t;\hat{\theta}) = G(t)^{\hat{\theta}}$ with a non-parametric estimator, for example the Nelson-Aalen estimator, $\bar{S}(t)$, in  (\ref{score_obs}) and (\ref{score_cen}). Since  $\bar{S}(t)$ and $S(t;\hat{\theta})$ will be asymptotically equivalent, the test statistic (\ref{peto_test}) will remain asymptotically efficient. This is exactly the log-rank statistic (since the data is considered fixed in a permutation test, the sum of scores on one arm only is an equivalent test statistic to taking the difference in mean scores -- see the Appendix).

Note how Peto \& Peto derive the log-rank scores very directly. Also in 1972, Cox \cite{cox1972regression} published his famous proportional hazards model, where, if treatment is the only covariate, the hazard function is $\lambda_0(t)$ on the control arm and $\lambda_0(t)\exp(\beta)$ on the experimental treatment arm. Inference proceeds via the partial likelihood for $\beta$,
\begin{equation*}
    L(\beta) = \Pi_{i=1}^n \left\lbrace \frac{\exp(\beta\times \mathbb{I}\left\lbrace \text{patient } i \text{ on trt }1 \right\rbrace)}{\sum_{l \in R(x_i)}\exp(\beta\times \mathbb{I}\left\lbrace \text{patient } l \text{ on trt }1 \right\rbrace)}\right\rbrace^{\delta_i},
\end{equation*}
where $\delta_i=1$ if patient $i=1,...,n$ has an observed event, and zero otherwise, where $x_i$ denotes the observation from patient $i$, and $R(x_i)$ denotes the risk set at time $x_i$. The efficient score statistic is
\begin{equation*}
    \left.\frac{\partial l(\beta)}{\partial \beta}\right\rvert_{\beta=0} = \sum_{j = 1}^{k} \left( O_{1,j} - E_{1,j} \right),
\end{equation*}
with $O_{1,j}$ and $E_{1,j}$ as defined above, and where the summation is now over the $k$ event times. Once again, this is exactly the log-rank statistic. Notice, however, that in contrast to  Peto \& Peto, what emerges from the Cox model is Mantel's "observed minus expected" form of the test statistic. 

\section{Weighted log-rank tests as permutation tests}

\subsection{Fleming-Harrington (0,1) test}

Given that the log-rank test can be viewed as both an observed-minus-expected test, and as a permutation test, it is natural to ask if the same is true for the Fleming-Harrington-(0,1) test. Leton \& Zuluaga \cite{leton2001equivalence} show that it is true for every weighted log-rank test, and they provide formulas for calculating the corresponding scores (see Appendix). Figure \ref{scores}(c) shows what the Fleming-Harrington-(0,1) scores look like when applied to our toy data set. Expressed in this way, the issues that were apparent in the simulation study are now easy to explain. The scores (for observed events) are no longer monotonically decreasing with time. This means, for example, that someone who dies at month 2 gets a lower, i.e., "better", score than someone who dies at month 17. This point bears repeating. We are claiming a success when the mean score on the experimental treatment is sufficiently lower than the mean score on control, yet the way that we assign scores to observations gives lower scores to earlier deaths than later deaths.

\subsection{Modestly-weighted log-rank test}

When expressed as a permutation-of-scores test, we have seen that the Fleming-Harrington-(0,1) test statistic may appear nonsensical, or at least inappropriate, for testing whether an experimental drug improves survival. Yet it clearly has high power in delayed-effect scenarios. Is it possible to construct an alternative form of weighted log-rank test that is powerful under delayed-effect scenarios but maintains a sensible interpretation when viewed as a permutation test? This was the motivation behind the modestly-weighted log-rank test (MWLRT) of Magirr \& Burman \cite{magirr2019modestly}. The idea, depicted in Figure \ref{scores}(d) for the toy example data, is to keep the scores fixed at $1$ for all events up to a pre-specified timepoint $t^*$. When viewed as a weighted observed-minus-expected test, this means that the weights start at around $1$ for the first event, and are increasing up until time $t^*$. Beyond time $t^*$, however, we choose to keep the  weights fixed at the largest pre-$t^*$ value. When viewed again as a permutation test, this means that the scores are decreasing beyond $t^*$. By using this procedure, we achieve the up-weighting of later event times, but also rule out the possibility of an earlier death being given a better score than a later death. It turns out that one can achieve this pattern by using a weight function:
\begin{equation}\label{modest_weight}
    w_j = 1 / \max \left\lbrace \hat{S}(t_{j}-), \hat{S}(t^*) \right\rbrace,
\end{equation}
where $\hat{S}(t)$ is the Kaplan-Meier estimate of the survival curve in the pooled data. This is highly convenient because this weight function fits into the family studied by Fleming and Harrington \cite[ch.~7]{fleming2011counting}, meaning that their asymptotic distributional results apply. Implementing the MWLRT is therefore simple. It is entirely analogous to implementing the Fleming-Harrington-(0,1) test. Note that (\ref{modest_weight}) uses $\hat{S}(t_{j}-)$ instead of the $\hat{S}(t_{j})$ used in \cite{magirr2019modestly}. This small change means that the scores are now only approximately equal to $1$ up to time $t^*$.

The MWLRT also has an appealing heuristic interpretation as a robust milestone (or landmark) analysis. A milestone analysis compares estimated survival curves at a pre-specified timepoint, $\tau$. This is a simple, intuitive, clinically-interpretable summary measure. For non-censored data, one could construct a corresponding permutation-of-scores test by, for example, assigning all survival times less than $\tau$ a score of $1$, and all survival times greater than or equal to $\tau$ a score of $-1$.
One reason, however, for not choosing a milestone as the primary analysis method is the difficulty in pre-specifying $\tau$, and the loss in power from getting this choice wrong. To protect against a bad choice of $\tau$, one could think about pre-specifying a range of potential milestones, $\tau_1,\dots,\tau_m$ and using an average
\begin{equation*}
    \frac{1}{m}\sum_{i=1}^m\hat{S}_1(\tau_i) - \hat{S}_0(\tau_i)
\end{equation*}
as the test statistic. The corresponding permutation-of-scores test would assign scores to observations using the average score across the $m$ individual milestone tests, which would be similar in shape to the MWLRT scores, i.e., fixed at 1 for a given period of time (up to $\tau_1$) before gradually declining. In this sense, the MWLRT can be thought of, heuristically, as similar to a robust milestone analysis. 

Finally, the MWLRT requires one to pre-specify a timepoint $t^*$. Given the multiple assumptions that go into designing a survival trial, including recruitment rates, event rates and follow-up times, one could in theory choose $t^*$ to optimize the power of the MWLRT (see \cite{magirr2019modestly}). However, as also shown in \cite{magirr2019modestly}, and in contrast to a milestone analysis, the power of the MWLRT is robust across a range of $t^*$. As a starting point, one could, for example, use the weight function 
\begin{equation*}
    w_j = 1 / \max \left\lbrace \hat{S}(t_{j}-), 0.5 \right\rbrace,
\end{equation*}
which means that the weights are increasing until the survival estimate in the pooled data drops to 0.5. This will often have good power in a delayed-effect scenario with reasonably mature data, for example, when there is enough follow-up for median survival times to be estimated from the Kaplan-Meier curves. Note that as $t^* \rightarrow \infty$, the MWLRT weights reduce to $w_j = 1 / \hat{S}(t_{j}-)$, which Gray \& Tsiatis \cite{gray1989linear} show are the optimal weights under a cure-rate model. As discussed by Gray \& Tsiatis, if the event rate is low, such that $\hat{S}(t_{j}-)$ is never too far from $1$, then there will be little difference between this test and a standard log-rank test.

\subsection{Simulation study (part 2)}

\begin{figure}
  \centering
  \includegraphics[width=\textwidth]{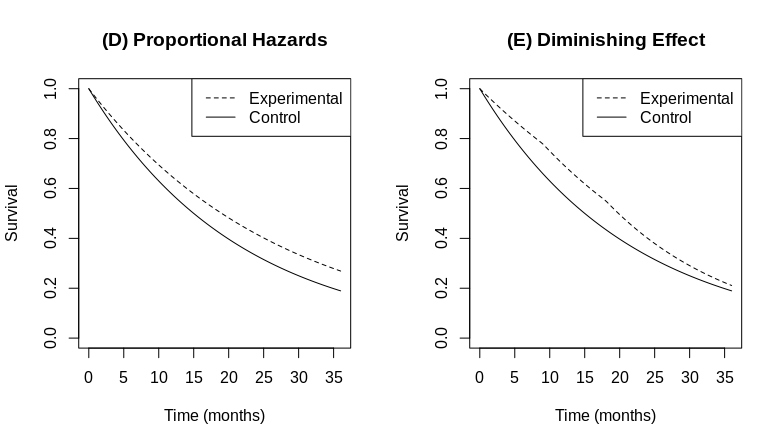}
  \caption{Additional scenarios used in the simulation study. (D) Control arm: exponential survival with median $15$ months. Experimental arm: exponential survival with median $19$ months. (E) Control arm: exponential survival with median $15$ months. Experimental arm: three-piece exponential with changepoints at $9$ months and $18$ months, with rates $\log(2)/25$, $\log(2)/18$ and $\log(2)/13$}
  \label{sim_scenarios_2}
\end{figure}

We return to our simulation study, where we consider a randomized controlled trial with 500 patients recruited to each treatment arm uniformly over a 12 month period, survival curves as shown in Figure \ref{sim_scenarios}, and administrative censoring 36 months after the start of the study. The augmented results in Table \ref{sim_2} show that under the delayed-effect scenario the MWLRT achieves much higher power than the standard log-rank test (although lower power than the Fleming-Harrington-(0,1) test). To put this difference into context, the relative efficiency (when $t^*=12$) is
\begin{equation*}
100\times\left\lbrace \frac{\Phi^{-1}(0.975) + \Phi^{-1}(0.89)}{\Phi^{-1}(0.975) + \Phi^{-1}(0.83)} \right\rbrace^2 = 120\%.
\end{equation*}
This means that to achieve the same power as the MWLRT, the sample size for the log-rank test would need to be increased by about 20\%. At the same time, the MWLRT  guarantees that the probability of claiming benefit when the experimental arm is uniformly worse than the control arm is below 2.5\%. With an astutely chosen timepoint ($\tau = 27$), a milestone analysis also has high power under the delayed-effect scenario, but one can see the risk of power loss if a suboptimal timepoint ($\tau = 21$) is chosen. In contrast, the MWLRT maintains good power across a wide range of $t^*$. To further elucidate the properties of the MWLRT, two further scenarios, shown in Figure \ref{sim_scenarios_2}, are added to the simulation study, with results also given in Table \ref{sim_2}. Under proportional hazards, the MWLRT maintains good power relative to the standard log-rank test. However, if the assumption of a delayed effect is completely wrong, and there is a strong initial effect that is diminishing over time, then the power of the MWLRT is much lower. 

\begin{table}[ht]
\caption{Results from the simulation study described in the text and Figures \ref{sim_scenarios} and \ref{sim_scenarios_2}. Based on 1000 repetitions}
\centering
\begin{tabular}{c|cccccc}
  & \multicolumn{6}{c}{Probability of claiming benefit} \\
Scenario & Log-rank & F-H-(0,1) & MWLRT(12) & MWLRT(24) & Milestone(21) & Milestone(27) \\ 
  \hline
(A) Delayed Effect & 0.83 & 0.93 & 0.89 & 0.91 & 0.78 & 0.87 \\ 
  (B) Identical & 0.02 & 0.03 & 0.02 & 0.02 & 0.02 & 0.03 \\ 
  (C) Worse than Control & 0.00 & 0.07 & 0.01 & 0.02 & 0.01 & 0.03 \\ 
  (D) Proportional Hazards & 0.89 & 0.78 & 0.88 & 0.86 & 0.78 & 0.83 \\ 
  (E) Diminishing Effect & 0.80 & 0.13 & 0.64 & 0.37 & 0.83 & 0.43 \\ 
   \hline
\end{tabular}
\label{sim_2}
\end{table}

\section{Discussion}

\subsection{From hypothesis testing to estimation}

Throughout this paper, we have framed the discussion around hypothesis testing. The null hypothesis that is often considered in survival analysis is $H_0: S_1(t) = S_0(t)$ for all $t$. However, we have seen that this is rather a narrow view. When testing an experimental drug in a confirmatory RCT, it is not sufficient to show that survival differs in some unspecified way to the control arm. At the very least, one must demonstrate that there is a survival benefit at at least some timepoints, and therefore $\tilde{H}_0: S_1(t) \leq S_0(t)$ for all $t$ is a more appropriate null hypothesis. But rejecting $\tilde{H}_0$ is only a minimal requirement for drug approval. Beyond this, one of course needs to estimate the treatment effect in order to judge its clinical relevance. In a well-designed confirmatory trial, serious forethought will have been given to this issue, such that formal rejection of $\tilde{H}_0$ will often coincide with an estimated effect size that is considered meaningful, as is discussed below. This should never be taken for granted, though. Dangerous situations include subgroup analysis \cite{dahlberg2020clinical}, and early analysis of overall survival when a trial has been powered for progression-free survival. And "meaningful effect" is of course subjective.

How is estimation typically handled? The predominant method is to plot Kaplan-Meier estimates of the survival curves on the two treatments arms. This is an excellent way to judge clinical relevance. The effect size is shown on an absolute scale, one sees the whole distribution, and it is unbiased in a single-stage (non-group-sequential) design provided that censoring is non-informative (sequential testing and issues with informative censoring are out of scope for this paper). It is not perfect, however, as it can be difficult to gauge the level of uncertainty surrounding the estimate, particularly in the tail of the distribution \cite{morris2019proposals}.

In addition to Kaplan-Meier estimates, the estimated hazard ratio parameter from a Cox model is usually reported, together with a confidence interval, regardless of whether the proportional hazards assumption appears reasonable or not. What does this number really mean? As shown by Berry et al. \cite{berry1991comparison}, the Peto hazard ratio estimate,
\begin{equation}\label{peto_hr}
    \hat{\theta} = \exp \left( \frac{\sum_{j}  O_{1,j} - E_{1,j}}{\sum_jV_{1,j}}\right),
\end{equation}
will be close to the Cox model estimate under certain (often reasonable) assumptions. One can express (\ref{peto_hr}) as a weighted geometric mean,
\begin{equation}\label{whr}
    \hat{\theta} = \exp \left( \frac{\sum_k  I_k \log\hat{\theta}_k}{\sum_k I_k} \right)
\end{equation}
where $\hat{\theta}_k$ is the estimate of the hazard ratio in a time interval $(\tau_{k-1},\tau_{k}]$ for some $0= \tau_0 < \tau_1 < \cdots < \tau_m$, and $I_k=\sum_{t_j\in(\tau_{k-1},\tau_{k}]}V_{1j}$  is approximately equal to the number of observed events in $(\tau_{k-1},\tau_{k}]$ divided by 4 (assuming 1:1 randomization). If the hazard ratio changes over time, the estimand targeted by $\hat{\theta}$ becomes a weighted average of the hazard ratio function with weights dictated by the number of observed events in each time period. In other words, the estimand is a function of both the time-to-event distributions and the censoring distributions, the latter being dictated chiefly by the recruitment rate and follow-up time \cite{xu2000estimating, rufibach2019treatment}. Strictly speaking, this is an absurd estimand for judging clinical relevance. Why should clinical relevance depend on the recruitment rate? In practice, of course, the hazard ratio parameter is used in a more limited sense. For example, an estimated hazard ratio of 0.5 with 95\% confidence interval (0.25,1) is associated with the same p-value as an estimated hazard ratio of 0.95 with 95\% confidence interval (0.9, 1). In the first case, we would say that we have a large relative effect that is imprecisely estimated, whereas in the second case we have a small, but precisely estimated, relative effect. This type of qualitative conclusion will often be valid even under non-proportional hazards (though perhaps not when one has an extreme case of crossing survival curves). Obviously, regardless of proportional hazards, the hazard ratio parameter says nothing about absolute treatment effects. A hazard ratio of 0.8 means something very different when median survival is 2 months compared to when it is 5 years. 

It should by now be clear that using the MWLRT instead of a standard log-rank test does not risk introducing a disconnect between hypothesis testing and interpretation of effect size. Such a disconnect already exists. And this is a good thing. Relying on the point estimate and confidence interval of the hazard ratio parameter from a Cox model would be inappropriate.

\subsection{Sample size considerations}

How do sample size calculations proceed when one anticipates a delayed-effect scenario? To give a flavour, one might start out by ignoring the delayed effect, and using very simple working assumptions: exponential survival distributions, a median survival of $\mu_0$ months on control, and a desire for high power should median survival on control be $\mu_1$ months. Assuming a standard log-rank statistic, for 1:1 randomization, the required number of events would be

\begin{equation*}
    n_e = 4\left( \frac{\Phi^{-1}(0.9) + \Phi^{-1}(0.975)}{-\log(\mu_0/\mu_1)} \right)^2,
\end{equation*}

assuming a one-sided significance level of 2.5\% and power 90\%, say. In this case, a hazard ratio point estimate of 

\begin{equation*}
    \hat{\theta}^* = \exp\left\lbrace \Phi^{-1}(0.025)\sqrt{\frac{4}{n_e}}\right\rbrace
\end{equation*}

would just about achieve statistical significance. Notice that there is a trade-off here. One needs large $n_e$ to have sufficient power, but making $n_e$ too large means $\hat{\theta}^*$ goes to $1$, i.e., there is a possibility of getting a significant but not clinically meaningful result. These two opposing forces should lead to a sensible $n_e$. The next step would be to convert the required number of events into a sample size, based on available knowledge regarding recruitment rates and maximum potential follow-up time. This step could be more or less sophisticated, but it is standard practice, and software is widely available. 

If the working assumption is considered implausible, and there is a strong belief that the hazard rates on the two arms will be similar for the first 3 or 4 months, for example, before diverging, then the next step would probably be to simulate the trial, based on the sample size derived from the working assumptions, but using the more realistic time-to-event distributions. The simulated power is likely to be less than 90\%, and one would need to tweak either the sample size or follow-up time to get it back up there. Perhaps a range of options would be considered. Note that this step is already a complex undertaking. If the MWLRT is used instead of the log-rank statistic, the only thing that changes is the test procedure used in the simulation study. The size of the task is essentially the same.
 
\subsection{Crossing survival curves}

If we lived in a world where survival distributions on the two arms were guaranteed to either belong to $\tilde{H}$, i.e., $S_1(t) \leq S_0(t)$ for all $t$, or to have $S_1(t) \geq S_0(t)$ for all $t$ (with strict superiority at least for some $t$), then there would perhaps be little controversy in choosing between a log-rank test, MWLRT, or a milestone as the primary analysis method, based purely on power considerations. Sometimes, however, the truth will be that $S_1(t) < S_0(t)$ for some $t$, and $S_1(t) > S_0(t)$ for other $t$. What do we want to happen in such situations? Do we want a test that is powerful in terms of picking out that $S_1(t) > S_0(t)$ for at least some $t$? Or do we see this as undesirable, and hence are looking for a conservative test that somehow balances out the different time periods to give a null result? 
This is impossible to answer in general, because it depends on how we trade off the value of short-term versus long-term survival. One thing that can be said, however, is that the standard log-rank test is by no means a neutral default. Contrary to what is sometimes claimed \cite{freidlin2019methods}, the standard log-rank statistic does not weight all parts of the survival curves equally. We have seen in (\ref{whr}) that it weights the survival curves based on how many events are observed in each time period. For example, with heavy early censoring, most events would happen in the early part of the curves, such that a fleeting early difference might be exaggerated. The point here is not that the standard log-rank test is a bad choice. Rather, no test will be uniformly most sensible under crossing survival curves. This underlines the importance of carefully assessing treatment effect size, in particular, via the Kaplan-Meier estimates.

\subsection{Concluding remarks}

In 2019, Freidlin \& Korn wrote a commentary in the Journal of Clinical Oncology entitled "Methods for accommodating non-proportional hazards in clinical trials: ready for the primary analysis?" Arguably, at least in the case of weighted log-rank statistics, the answer to that question was indeed "no". When viewed as a permutation-of-scores test, the Fleming-Harrington-(0,1) statistic assigns better scores to worse outcomes. In the context of the primary analysis of a confirmatory clinical trial testing whether experimental treatment improves survival, this is a fundamental weakness. Many recently proposed methods, for example the "max-combo" test \cite{lin2020alternative, roychoudhury2019robust}, incorporate the Fleming-Harrington-(0,1) test as a component, and therefore inherit this weakness. Fortunately, however, the permutation-of-scores perspective also shows us how to rectify this issue. The MWLRT has the key property that if survival on the experimental drug is truly lower (or equal) to survival on control at all timepoints, then the probability of claiming a statistically significant improvement is less than 2.5\% (assuming a conventional threshold is applied). In addition, for many common scenarios in immuno-oncology, sample size can be reduced by 10 -- 20\%, compared to using a standard log-rank test, without losing power. On top of power and false-positive-rate considerations, it is also very simple to implement, comes with rigorous asymptotic results, and has an appealing heuristic interpretation as a robust milestone analysis. It therefore deserves serious consideration as the primary analysis method in confirmatory trials. Interpretation of treatment effect size does not come for free. But it never does in survival analysis. 

\section*{Appendix}

Following Leton \& Zuluaga \cite{leton2001equivalence}, and letting $l_{1,j}$ and $l_{0,j}$ denote the number of patients censored on the test treatment and control treatment, respectively, during $\left[\left.t_j, t_{j+1}\right)\right.$,

\begin{align*}
U &:=\sum_{j = 1}^{k} \left( O_{1,j} - O_j\frac{n_{1,j}}{n_j} \right)\\
  &=\sum_{j = 1}^{k} O_{1,j} - \sum_{j = 1}^{k}\frac{O_{j}}{n_j} \times n_{1,j}\\
  &=\sum_{j = 1}^{k} O_{1,j} - \sum_{j = 1}^{k}\frac{O_{j}}{n_j} \times \sum_{i = j}^{k}(O_{1,i} + l_{1,i})\\
  &=\sum_{j = 1}^{k} O_{1,j} - \sum_{j = 1}^{k}(O_{1,j} + l_{1,j}) \times \sum_{i = 1}^{j}\frac{O_{i}}{n_i}\\
  &=\sum_{j = 1}^{k} O_{1,j}\left( 1 - \sum_{i = 1}^{j}\frac{O_{i}}{n_i} \right) +  \sum_{j = 1}^{k}l_{1,j} \left( - \sum_{i = 1}^{j}\frac{O_{i}}{n_i}\right) \\
  &=\sum_{j = 1}^{k} O_{1,j} \times \left\lbrace 1 + \log(\bar{S}_{NA}(t_j))\right\rbrace +  \sum_{j = 1}^{k} l_{1,j}\times \log(\bar{S}_{NA}(t_j)).
\end{align*}

Ranking the observations, regardless of censoring, $x_{(n)}\leq x_{(n-1)} \leq \cdots \leq x_{(2)} \leq x_{(1)},$ and denoting the corresponding treatment assignments as, $z_{(n)}\leq z_{(n-1)} \leq \cdots \leq z_{(2)} \leq z_{(1)}$, one can write $U = \sum_{i= 1}^n \mathbb{I}\left\lbrace z_{(i)} = 1 \right\rbrace a_{(i)},$ where 
\begin{equation*}
    a_{(i)} = \mathbb{I}\left\lbrace x_{(i)} \text{ not censored} \right\rbrace + \log(\bar{S}_{NA}(x_{(i)})).
\end{equation*}
Therefore one can perform the log-rank test by permuting treatment labels very many times, each time recalculating $U$, and counting the proportion of times that the it is less than or equal to the original value. Note, however, that instead of using $U$ one could also use
\begin{equation*}
    \tilde{U} = \frac{\sum_{i= 1}^n \mathbb{I}\left\lbrace z_{(i)} = 1 \right\rbrace a_{(i)}}{\sum_{i= 1}^n \mathbb{I}\left\lbrace z_{(i)} = 1 \right\rbrace} - \frac{\sum_{i= 1}^n \mathbb{I}\left\lbrace z_{(i)} = 0 \right\rbrace a_{(i)}}{\sum_{i= 1}^n \mathbb{I}\left\lbrace z_{(i)} = 0 \right\rbrace},
\end{equation*}
as the test statistic in the permutation test, as $U$ and $\tilde{U}$ are equivalent up to a (positive) scale and shift transformation, i.e.,
\begin{equation*}
    \tilde{U} = \frac{n}{\sum_{i= 1}^n \mathbb{I}\left\lbrace z_{(i)} = 1 \right\rbrace\sum_{i= 1}^n \mathbb{I}\left\lbrace z_{(i)} = 0 \right\rbrace} \times U - \frac{\sum_{i= 1}^n  a_{(i)}  \sum_{i= 1}^n \mathbb{I}\left\lbrace z_{(i)} = 1 \right\rbrace}{\sum_{i= 1}^n \mathbb{I}\left\lbrace z_{(i)} = 1 \right\rbrace\sum_{i= 1}^n \mathbb{I}\left\lbrace z_{(i)} = 0 \right\rbrace}.
\end{equation*}

In general, when we introduce weights:

\begin{align*}
U_W &:=\sum_{j = 1}^{k} w_j\left( O_{1,j} - O_j\frac{n_{1,j}}{n_j} \right)\\
  &=\sum_{j = 1}^{k} w_jO_{1,j} - \sum_{j = 1}^{k}w_j\frac{O_{j}}{n_j} \times n_{1,j}\\
  &=\sum_{j = 1}^{k} w_jO_{1,j} - \sum_{j = 1}^{k}w_j\frac{O_{j}}{n_j} \times \sum_{i = j}^{k}(O_{1,i} + l_{1,i})\\
  &=\sum_{j = 1}^{k} w_jO_{1,j} - \sum_{j = 1}^{k}(O_{1,j} + l_{1,j}) \times \sum_{i = 1}^{j}w_i\frac{O_{i}}{n_i}\\
  &=\sum_{j = 1}^{k} O_{1,j}\left( w_j - \sum_{i = 1}^{j}w_i\frac{O_{i}}{n_i} \right) +  \sum_{j = 1}^{k}l_{1,j} \left( - \sum_{i = 1}^{j}w_i\frac{O_{i}}{n_i}\right).
\end{align*}

This means that an observed event at time $t_j$ is given a score of $a = w_j - \sum_{i = 1}^{j}w_i\frac{O_{i}}{n_i}$, and an observation censored during $\left[\left.t_j, t_{j+1}\right)\right.$ is given a score of $a = - \sum_{i = 1}^{j}w_i\frac{O_{i}}{n_i}$.

\bibliographystyle{unsrt}  
\bibliography{references} 
\end{document}